\documentclass[a4paper]{article}

\usepackage{icrc2013}

\newcommand{\etal}{et al.}

\title{Systematic search for $\gamma$-ray emitting molecular clouds in the vicinity of supernova remnants}

\shorttitle{Search for $\gamma$-ray emitting molecular clouds}

\authors{
Stephanie H\"affner$^{1,2}$,
Ira Jung$^{1}$,
Christian Stegmann$^{3,2}$,
}

\afiliations{
$^1$ Erlangen Centre for Astroparticle Physics (ECAP), Universit\"at Erlangen-N\"urnberg, Erwin-Rommel-Str. 1, D-91058 Erlangen, Germany \\
$^2$ Institut f\"ur Physik und Astronomie,  Universit\"at Potsdam, Karl-Liebknecht-Str. 24/25, D-14476 Potsdam-Golm, Germany \\
$^3$ DESY, Platanenalle 6, D-15738 Zeuthen, Germany \\
}

\email{stephanie.haeffner@fau.de}

\abstract{Observations of very-high-energy (VHE) $\gamma$-ray emission from supernova remnants (SNR) established them as sources of accelerated particles up to energies of 100$\,$TeV. The dominant process $\--$ leptonic or hadronic $\--$ responsible for the VHE emission is still not proven for most of the SNRs.

Molecular clouds (MCs) in the vicinity of SNRs provide increased amount of target material for accelerated particles escaping the SNRs, thus making MCs potential $\gamma$-ray sources.
The predicted $\gamma$-ray flux for MCs offset from the SNR shock depends on the applied diffusion model for VHE particles and the SNR and MC properties, which encounter large uncertainties. While the the average galactic diffusion coefficient is estimated, the spatially resolved propagation properties of VHE cosmic rays are unknown. 
$\gamma$-ray emitting MCs provide a unique possibility to derive new information on the propagation of VHE particles through the ISM and on the acceleration of hadrons at SNRs.

We present in this paper a strategy and first results of a systematic search for TeV $\gamma$-ray emitting MCs in the vicinity of SNRs in the galactic plane region. Based on a theoretical model and current SNR and MC catalogues the detectable parameter space (e.g. diffusion coefficient, SNR age) for current Imaging Atmospheric Cherenkov Telescope systems (IACTs) is obtained. This allows to identify potential $\gamma$-ray emitting regions.}


\begin{document}
\maketitle

\section{Introduction}

Supernova remnants (SNRs) are prime candidates for Galactic sources of Cosmic Rays (CRs) up to energies of $10^{15}\,$eV. Observations of SNRs in the very-high-energy (VHE) $\gamma$-ray domain support this assumption, but for most of the SNRs it is not clear whether the observed $\gamma$-ray emission is produced by accelerated leptons via Inverse Compton scattering or by accelerated hadrons interacting with the ambient medium and producing $\gamma$-rays via proton-proton collision and subsequent pion decay. Recently,  the \textit{Fermi}-LAT Collaboration reported the detection of pion-decay signatures in the SNRs W44 and IC 443 \cite{Ackermann2013}. The detection of $\gamma$-ray emission outside the SNR shell coincident with ambient Molecular Clouds (MCs) strengthens these results \cite{Uchiyama2012}. 
In this scenario hadrons were accelerated in the SNR shock and afterwards
escaped the acceleration site. These high-energy particles can interact then
with MCs in the vicinity of the SNR. MCs provide an increased amount of target material
and $\gamma$-ray emission via the neutral pion decay occurs.

Besides the possibility to probe hadronic acceleration in SNRs, $\gamma$-ray emitting MCs close to SNRs are also good laboratories to study propagation properties of VHE particles through the interstellar medium (ISM) from the SNR to the MC.
The detection of TeV $\gamma$-ray emission of the W28 region coincident with MCs was
reported by the H.E.S.S. Collaboration \cite{HESS2008}. The emission is visible in three distinct regions, two coincident with MCs offset from the SNR, one coincident with a MC in
the shock region.  The emission at the shock and at one position offset is also observed in the GeV energy regime by the \textit{Fermi}-LAT Collaboration \cite{Abdo2010}. The W28 region has been used to constrain the propagation properties of the VHE particles. The application of isotropic diffusion models to that region \cite{Gabici2010,Li2010,Ohira2011,Yan2012} shows that the diffusion at 10$\,$GeV is suppressed by more than one order of magnitude compared to the Galactic average value of $\approx 10^{28}\,$cm$^{2}\,$s$^{-1}$ \cite{Strong2007}.

Numerical simulations of the propagation of VHE particles \cite{Giacinti2012} predict
anisotropic diffusion with some filamentary structures. 
The application of an anisotropic diffusion model on the W28 region \cite{Nava2013} resulted in a diffusion coefficient in the order of $10^{28}\,$cm$^{2\,}$s$^{-1}$, whereas an isotropic diffusion coefficient is one and a half orders of magnitude smaller. To get further insights into the propagation of VHE particles studies of further MC-SNR associations are needed. 

We present in this paper a search for TeV $\gamma$-ray emitting MCs, which have an offset
to the SNR. This provides a unique possibility to test for the acceleration of hadrons
in SNRs and investigate local diffusion properties of VHE particles.
In Section \ref{sec:Gammaemission} we discuss the dependencies of the $\gamma$-ray emission of various SNR and MC  properties and of the diffusion coefficient as well as the detectability with current IACTs.
In Section \ref{sec:search}, the strategy for a general search for TeV $\gamma$-ray emitting MC-SNR associations is described and the application to the Galactic Ring Survey
(GRS) region is presented. A short discussion of the W44 region that is located within
the region covered by the GRS is given in the end. A conclusion is given in Section \ref{sec:conclusion}.

\section{$\gamma$-ray emission from MCs near SNRs}\label{sec:Gammaemission}

Different theoretical models exist to determine  the expected $\gamma$-ray emission
from MCs that are illuminated by escaping particles from a nearby SNR. These
models range from simple approaches to elaborated models using numerical simulations. 

To ensure a general search applicable to a broad SNR and MC population for the calculation of TeV $\gamma$-ray emission the isotropic model
from \cite{Gabici2009} is used to estimate the proton density at the location of the MC and \cite{Kelner2006} to calculate the expected $\gamma$-ray emission from pion decay due to hadronic interactions. 

\begin{figure}[htb]
  \centering
  \includegraphics[width=0.5\textwidth]{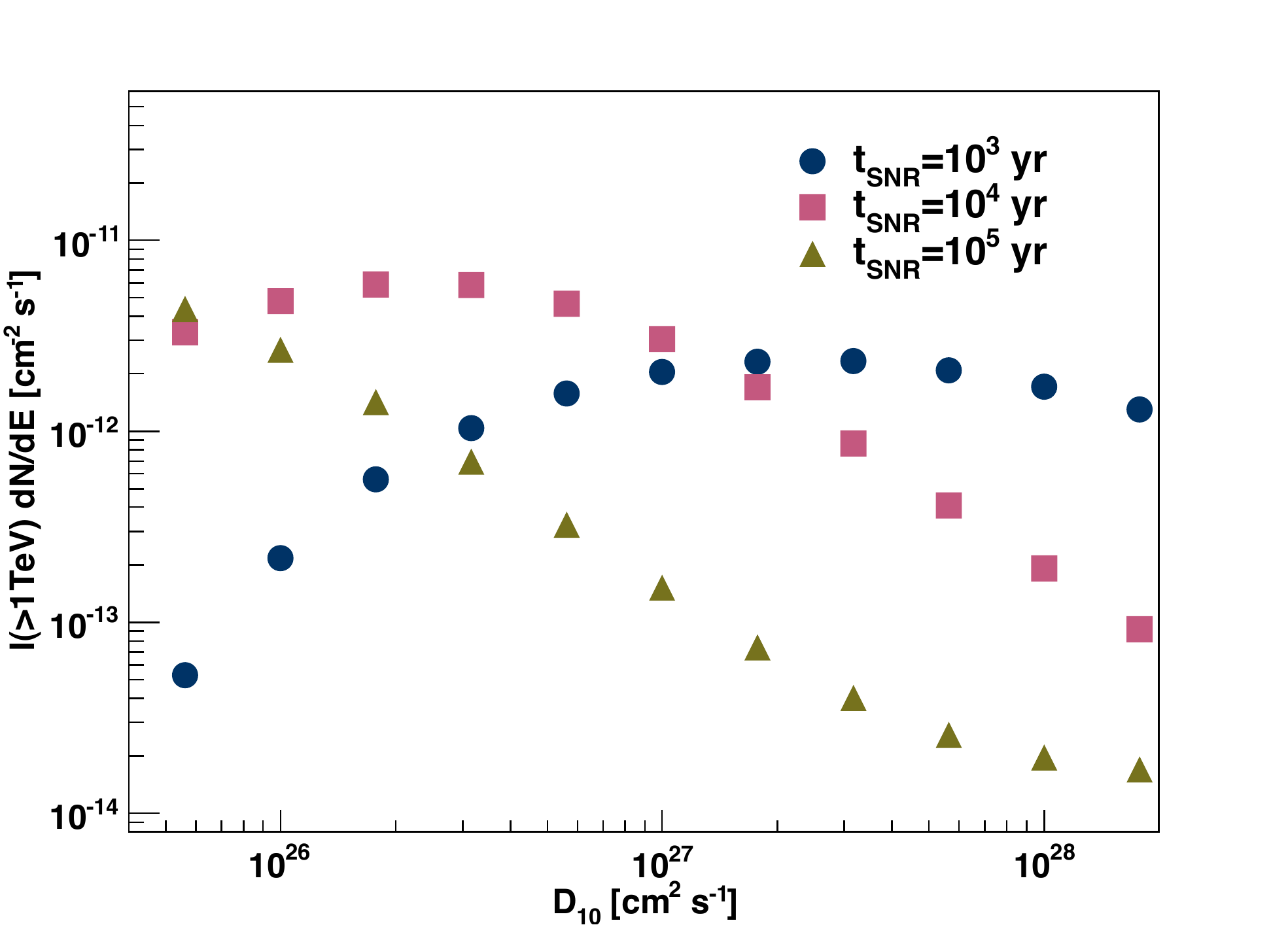}
  \caption{Integrated $\gamma$-ray flux above 1$\,$TeV as a function of D$_{10}$ and three assumed SNR ages. The SNR is situated at a distance of 50$\,$pc from the MC with a mass of $4\cdot 10^{4}\,$M$_{\odot}$ and a distance to Earth of 2$\,$kpc.}
  \label{fig:1}
 \end{figure}

Figure \ref{fig:1} shows integrated $\gamma$-ray fluxes above 1$\,$TeV of a MC offset from a SNR as a function of the normalisation of the diffusion coefficient $D_{10}$ at 10 GeV for three
different ages of a SNR. The energy dependence of the diffusion coefficient is considered to be of the form  $D(E)=D_{10}\cdot(\frac{E}{10\,\mathrm{GeV}})^s$, with $s=0.5$. These $\gamma$-ray fluxes are estimated for a MC with mass of $4 \cdot 10^{4}\,$M$_{\odot}$  at a distance of 50$\,$pc to the SNR and 2$\,$kpc to Earth. The MC mass and distance to Earth for this example resemble the properties of the MC properties as in the $\gamma$-ray emission scenario on the W28 region \cite{HESS2008}.
We assume that the Sedov time is 200$\,$yr. This is the time until the SNR enters
the Sedov phase in which the particle acceleration takes place. 30$\%$ of the SNR explosion energy of $10^{51}\,$erg is converted into the
acceleration of CRs. 

\begin{figure}[tb]
  \centering
  \includegraphics[width=0.5\textwidth]{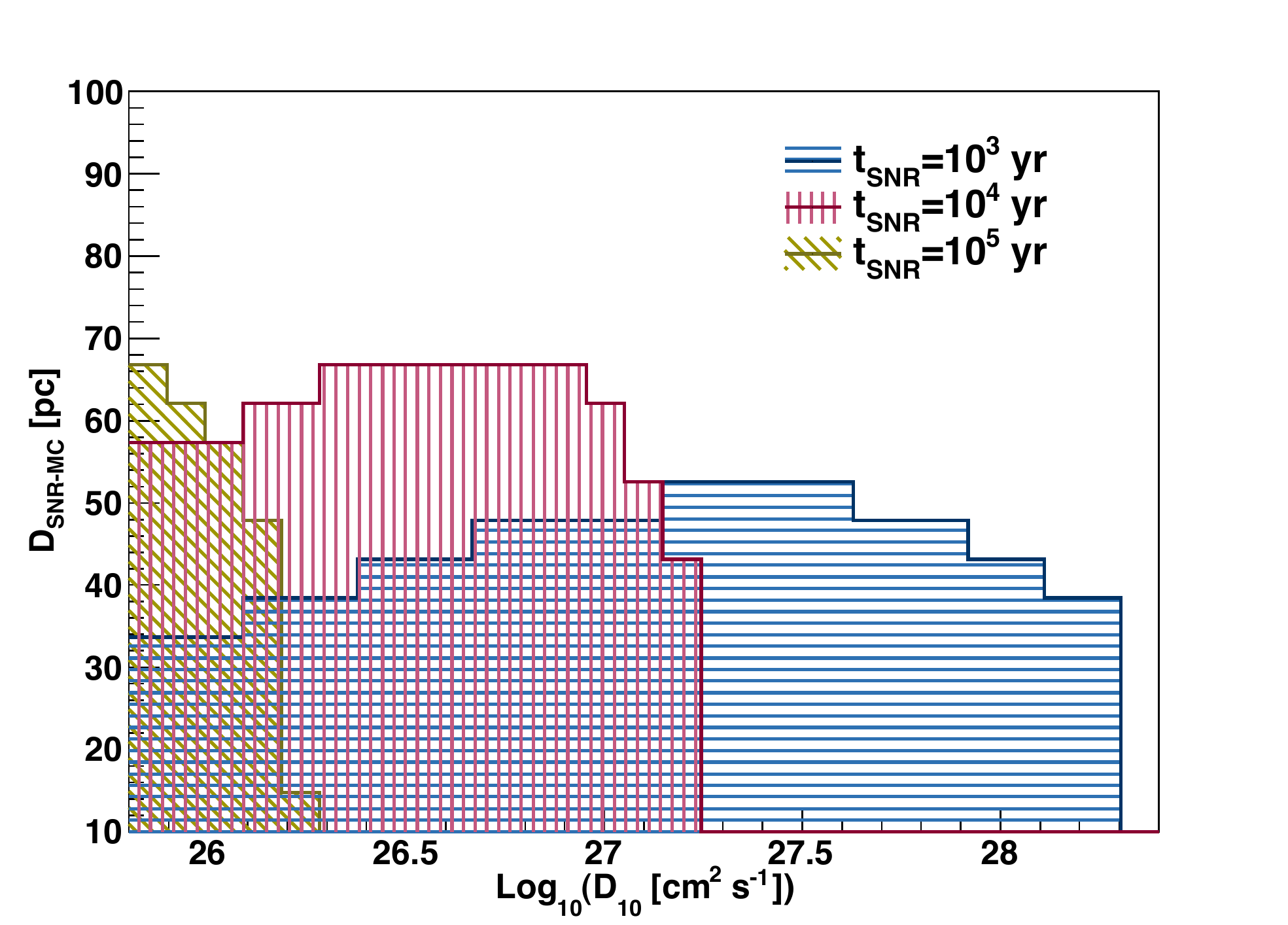}
  \caption{The detectable (I$(>$1$\,$TeV$)>4.5 \cdot 10^{-13\,}$cm$^{-2} \,$s$^{-1}$) parameter space for the diffusion coefficient and distance between SNR and MC is shown exemplarily for a MC with a mass of $4\cdot 10^{4}\,$M$_{\odot}$ and a distance to Earth of 2$\,$kpc.}
  \label{fig:2}
 \end{figure}

Current IACTs reach integrated sensitivities above 1$\,$TeV in the order of $1-2\%\,$Crab units (C.U.)\footnote{here: 1$\,$C.U.$\,$=$\,2.26\cdot 10^{-11}\,$cm$^{-2}\,$s$^{-1}$} (see e.g. the H.E.S.S. sensitivity for large parts of the  Galactic plane  \cite{Gast2011}). The comparison with the derived TeV $\gamma$-ray fluxes for the example scenario in Fig. \ref{fig:1} shows that a wide range of diffusion coefficients and SNR ages can give detectable $\gamma$-ray fluxes. The possible parameter space for the detection of  SNR-MC association in $\gamma$-rays depends on the diffusion coefficient, SNR ages and on the distance $D_{\mathrm{SNR-MC}}$ between the SNR and the MC is shown in Fig. \ref{fig:2}. The filled areas give the distances between SNR and MC in dependency of $D_{10}$  for three different SNR ages that lead for the example given above to an integrated $\gamma$-ray flux larger than $4.5 \cdot 10^{-13\,}$cm$^{-2} \,$s$^{-1}$. This value refers to a point-like sensitivity of 1$\%\,$C.U., linearly scaled for a cloud extent of $0.2^{\circ}$ in radius. MCs near young SNRs would be detectable in TeV $\gamma$-rays for a large range of diffusion coefficients and a distance smaller than 70 pc whereas for old SNRs ambient MCs would just be detectable for low diffusion coefficients $D_{10}\approx10^{26}\,$cm$^{2}\,$s$^{-1}$.
 
\section{Search for TeV $\gamma$-ray emitting MCs near SNRs}\label{sec:search}
\subsection{Strategy and application to the GRS region}
For the search of TeV $\gamma$-ray emitting MCs in the vicinity of SNRs we need the properties of SNRs and MCs to calculate the expected $\gamma$-ray flux. The detailed analysis of the TeV $\gamma$-ray data afterwards allows to constrain the parameter space and the model assumptions.

In this paper the isotropic model of \cite{Gabici2009} is used for the calculation of TeV $\gamma$-ray fluxes,  because more elaborated models that rely on detailed information of the individual SNR environments, e.g. densities and magnetic field orientation, are not suited for a general search. The input parameters of the theoretical model \cite{Gabici2009} are the following: the age of the SNR $t_{\mathrm{SNR}}$, the distance $D_{\mathrm{SNR-MC}}$ between the SNR and the MC, the MC mass $M_{\mathrm{MC}}$ and distance to the Earth $D_{\mathrm{MC}}$, the energy of the SNR explosion that was converted into CR acceleration $\eta \cdot E_{SN}$, the normalisation $D_{10}$ and energy dependence $s$ of the diffusion coefficient. Although not all of these parameters are well known, a general search for interesting regions of MC emitting detectable TeV $\gamma$-ray emission is possible with our strategy as described below.

MC and SNR properties from published catalogues are used for the search. The Galactic Ring Survey (GRS) is a survey of $^{13}$CO emission that covers the region between 18$^{\circ}$ and 56$^{\circ}$ Galactic longitude and between -1$^{\circ}$ and +1$^{\circ}$ Galactic latitude \cite{Jackson2006}. MCs and clumps have been identified within these data \cite{Rathborne2009}, the distance ambiguity was solved \cite{Roman-Duval2009} and the masses of the identified clouds have been determined \cite{Roman-Duval2010}. We selected 16 SNRs with measured ages and distances from \cite{Ferrand2012} that are located within the region of the GRS. The SNR ages range from 720$\,$yr to $10^5\,$yr, the SNR distances from $2.3\,$kpc to $11\,$kpc.

To take into account the uncertainties of the different parameters a parameter
scan was performed. The scan ranges and the step sizes are given in Table\ref{table_1}.

\begin{table}[h]
\begin{center}
\caption{Values and steps of the parameter scan for the calculation of $\gamma$-ray fluxes for each SNR-MC pair.}
\begin{tabular}{|l|c|c|}
\hline Parameter & Value & Steps \\ \hline\hline
$D_{10}$   & [$10^{26}$... $10^{28}]\,$cm$^{2}\,$s$^{-1}$  & [$10^{26}$,$10^{27}$,$10^{28}]$\\ 
 & & cm$^{2}\,$s$^{-1}$\\ \hline
$s$   &0.5  & fixed \\ \hline
$\eta\cdot E_{SN}$ & $0.3\cdot10^{51}\,$erg  & fixed \\ \hline
$t_{SNR}$ & $t_{min} \--t_{max}$  & 5 different ages \\ \hline
$d_{SNR}$ & $d_{min} \--d_{max}$  & 15$\,$pc steps \\ \hline
$M_{MC}$ & $M_{MC} \pm \Delta M_{MC}$  & 3 different masses \\ \hline
$d_{MC}$ & $d_{MC} \pm \Delta d_{MC}$  & 15$\,$pc steps \\ \hline

\end{tabular}
\label{table_1}
\end{center}
\end{table}

As mentioned above the local diffusion coefficient and its energy dependence in general are not known, just the average Galactic normalisation is determined and is of the order of $10^{28}\,$cm$^{2}\,$s$^{-1}$ \cite{Strong2007}. We follow the usual approach (see e.g. \cite{Pedaletti2013}) and assume the range between  $10^{26}\,$cm$^{2}\,$s$^{-1}$ and $10^{28}\,$cm$^{2}\,$s$^{-1}$ for $D_{10}$ and a power law with index $s=0.5$ for the energy dependence as suitable values. The minimum and maximum values for the SNR ages and distances are taken from \cite{Ferrand2012}. If no errors or uncertainties are listed in the catalogue, the given value is used without errors. The range of the SNR age is scanned with 5 steps. Considering the large uncertainty of $D_{10}$ this is a good compromise.
The errors on the MC distances are estimated in \cite{Roman-Duval2009} to be of the order of 20$\,\%$-30$\,\%$, we use therefore as distance uncertainty $\Delta d_{MC}=0.2\cdot d_{mc}$. The distance between the SNR and the MC is defined as the distance between the
centre position of the SNR and the peak of highest emission of the MC. The mean radius of the MCs is $24.1\,$pc \cite{Rathborne2009}, that is small compared to the errors on the MC and SNR distances. The errors on the MC masses $\Delta M_{MC}$ are given in \cite{Roman-Duval2010} and vary from about 10$\,\%$ to more than 40$\,\%$. The MC mass is a normalisation factor for the calculated $\gamma$-ray flux and three different values $M_{MC}$ and  $M_{MC}\pm\Delta M_{MC}$ are used in the parameter scan.

\subsection{Results and discussion}
The expected TeV $\gamma$-ray emission of MC-SNR associations was calculated for the
parameter ranges given above and then compared to the sensitivity of the current
generation of IACTs. The sensitivity of the H.E.S.S. Galactic Plane Scan \cite{Gast2011} that includes the region covered by the GRS was taken. About 200 out of the 749 MCs from the GRS sample have at least one parameter set that gives an integrated flux above the H.E.S.S. sensitivity\footnote{The pointlike sensitivity is multiplied by $\theta_{source}/0.1^{\circ}$ to take the MC extent into account.}. This large number of potential $\gamma$-ray emitters mainly results from the large uncertainties for the MC and SNR distances.
The clouds with at least one parameter set that gives a detectable flux are not distributed uniformly in the Galactic plane, but are clustered in certain regions around the SNRs, which allows to identify interesting regions.
More details can be found in our forthcoming paper. 

 \subsection{The vicinity of W44}\label{sec:W44}
 A prominent SNR with ambient MCs within the GRS region is the SNR W44 and its vicinity. W44 is surrounded by a complex of giant molecular clouds (e.g. \cite{Seta1994}). In the $\gamma$-ray regime the SNR shell was detected by EGRET (as 2EG J1857+0018) \cite{Thompson1995}, \textit{AGILE} \cite{Guliani2011} and also by \textit{Fermi}-LAT \cite{Abdo2010}. In addition, two regions close to the shell were also detected by \textit{Fermi}-LAT \cite{Uchiyama2012}. They coincide with ambient molecular clouds and the $\gamma$-ray emission is interpreted as a hint for hadron acceleration. This argument is strengthened by the pion-decay signature, which was recently reported by the \textit{Fermi}-LAT Collaboration \cite{Ackermann2013}.
 
\begin{figure}[htbp]
  \centering
  \includegraphics[width=0.5\textwidth]{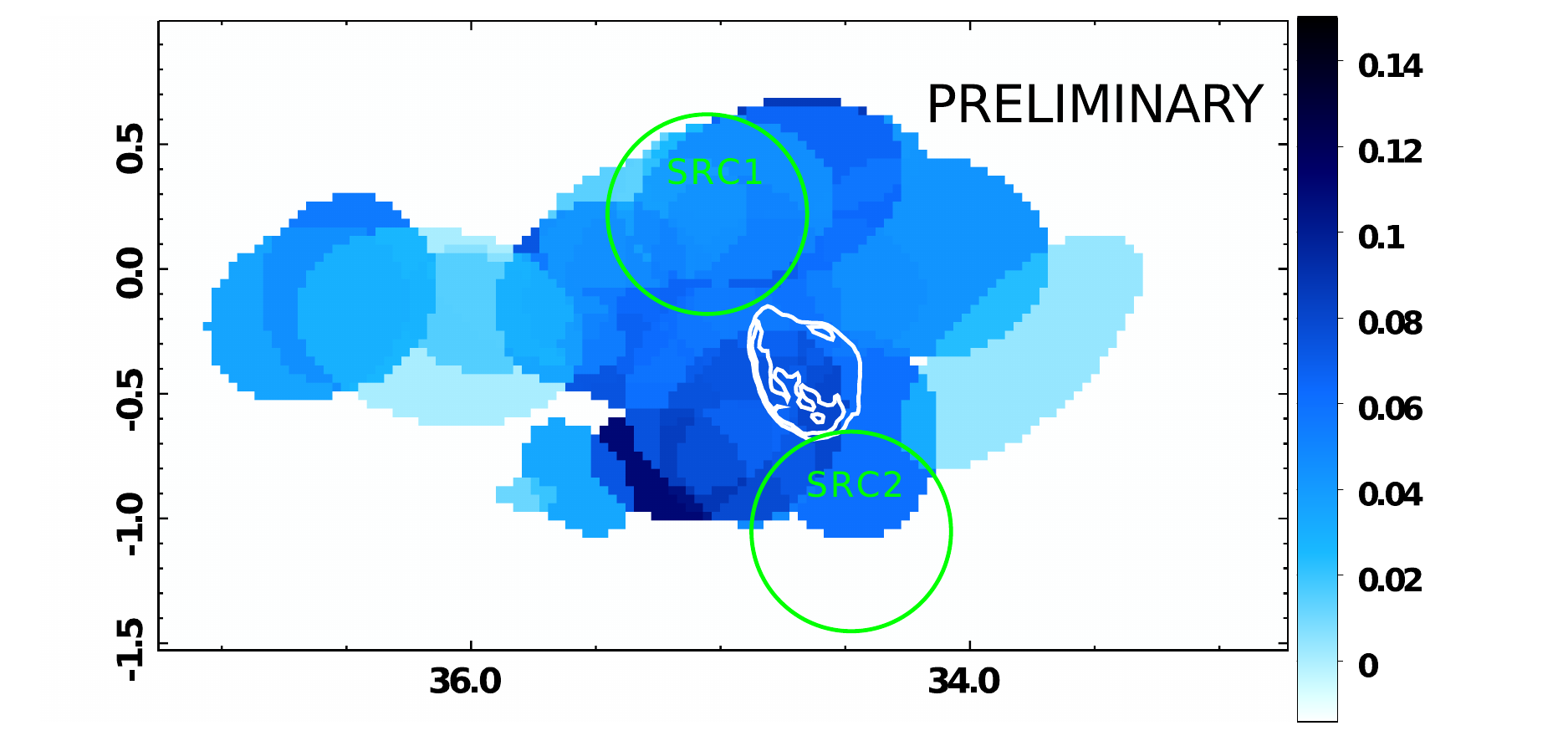}
  \caption{The region around the SNR W44. The radio contours of the SNR W44 are overlaid in white, the regions $SRC\,1$ and $SRC\,2$, identified by \cite{Uchiyama2012}, are marked by the green circles. The colour scale of the elipses, representing the MCs, gives the ratio of the number of parameter sets with fluxes above the H.E.S.S. Galactic Plane Scan sensitivity and the total number of parameter sets tested for the MC. }
  \label{fig:4}
 \end{figure}

Figure \ref{fig:4} shows a sky map of the W44 region. The ellipses represent the GRS MCs identified in \cite{Rathborne2009} and the colour scale gives the ratio of the number of parameter sets with fluxes above the H.E.S.S. Galactic Plane Scan sensitivity and the total number of parameter sets tested for the MC. The SNR radio shell is overlaid in white and the GeV $\gamma$-ray emitting regions outside the shell $SRC\,1$ and $SRC\,2$ as reported in \cite{Uchiyama2012} are overlaid in green.

In the following we will concentrate on the region $SRC\,1$, because the location of $SRC\,2$ is not fully covered by the GRS. We take all MCs from our search that lie partially or fully within the $SRC\,1$ region and sum up the individual predicted $\gamma$-ray emission with fixed MC and SNR distance for three different ages, 6$\,$000$\,$yr,  20$\,$000$\,$yr and  29$\,$000$\,$yr. The authors of \cite{Ferrand2012} give an age between 6$\,$000$\,$yr and 29$\,$000$\,$yr for W44, whereas in most publications an age of 20$\,$000$\,$yr is assumed. A value of 129$\,$yr \cite{Uchiyama2012} for the Sedov time and $0.3\cdot10^{51}\,$erg as fraction of the supernova explosion converted into CR acceleration was used. This is compatible with $0.3-3\cdot 10^{50}\,$ erg of energy channelled into the escaping CRs estimated by \cite{Uchiyama2012}.

Diffusion coefficients smaller than $2.5\cdot 10^{27}\,$cm$^{2}\,$s$^{-1}$, $6.3\cdot 10^{26}\,$cm$^{2}\,$s$^{-1}$ and $4.0\cdot 10^{26}\,$cm$^{2}\,$s$^{-1}$ for an age of 6$\,$000$\,$yr,  20$\,$000$\,$yr and  29$\,$000$\,$yr, respectively, give fluxes larger than the sensitivity of H.E.S.S. for the W44 region.
These results do not take into account the uncertainties of the SNR and MC properties. The emission of 4 MC contributes to the total expected emission of the $SRC\,1$ region. If one considers the uncertainties of the MC distances and masses a total of more than 10 MCs may contribute.


We conclude for the W44 region, that with next generation IACTs the detection of $\gamma$-ray emission is expected based on the GeV observations of the  $\gamma$-ray emitting MCs outside the shell and the extrapolation of the modelled $\gamma$-ray emission from \cite{Uchiyama2012} to higher energies. In case of the complex MC surrounding it is difficult to model $\gamma$-ray emission of the individual clouds because of the large distance uncertainties and the large number of clouds that could contribute in principal to the $\gamma$-ray emission. Therefore for diffusion studies regions with simpler environment are better suited.
  
\section{Conclusions}\label{sec:conclusion}
Molecular clouds near supernova remnants are important regions to study hadronic acceleration in SNR shocks and also diffusion properties of VHE particles. We presented our search strategy and discussed the caveats. The largest problems are the uncertainties on the SNR and MC distances. But nevertheless some promising regions can be identified. The W44 region is a  promising region to be detected with next generation IACTs, but for diffusion studies simpler environments are better suited.



\begin{thebibliography}{}

\bibitem{Ackermann2013} M. Ackermann, \etal, Science 339 (2013) 807-811

\bibitem{Uchiyama2012} Y. Uchiyama, \etal, ApJ 749 (2012) L35

\bibitem{HESS2008} H.E.S.S. Collaboration, F.A. Aharonian, \etal, A\&A  481 (2008) 401

\bibitem{Abdo2010} A.A. Abdo, \etal, ApJ 718 (2010) 348

\bibitem{Gabici2010} S.Gabici, S. Casanova, F.A. Aharonian,G. Rowell, in S.Boissier, M. Heydari-Malayeri, R. Samadi, D. Valls-Gabaud, eds, SF2A-2010: Proc. Annu. Meeting French Soc. Astron. Astrophys. Marseilles (2010), p. 313 (arXiv:1009.5291)

\bibitem{Li2010} H. Li and Y. Chen, MNRAS 409 (2010) L35

\bibitem{Ohira2011} Y. Ohira, K. Murase and R. Yamazaki, MNRAS 410 (2011) 1577

\bibitem{Yan2012} H. Yan, A. Lazarian and R. Schlickeiser, ApJ 745 (2012) 140

\bibitem{Strong2007} A.W. Strong, I.V. Moskalenko and V.S. Ptuskin, Annu. Rev. Nucl. Part. Sci. 57 (2007) 285-327

\bibitem{Giacinti2012} G. Giacinti, M. Kachelrie\ss ~and D.V. Semikoz, Phys. Rev. Lett. 108 (2012) 261101

\bibitem{Nava2013} L. Nava and S. Gabici, MNRAS 429 (2013) 1643-1651 

\bibitem{Gabici2009} S. Gabici, F.A. Aharonian and S. Casanova, MNRAS 396 (2009) 1629

\bibitem{Kelner2006} S.R. Kelner, F.A. Aharonian and V.V. Bugayov, Phys. Rev. D 74 (2006) 034018

\bibitem{Gast2011} H. Gast, \etal, for the H.E.S.S. Coll., ICRC proceeding (2011), arXiv:1204.5860

\bibitem{Jackson2006}  J.M. Jackson, \etal, ApJS 163 (2006) 145

\bibitem{Rathborne2009} J.M. Rathborne, \etal, ApJS 182 (2009) 131-142 

\bibitem{Roman-Duval2009} J. Roman-Duval, \etal, ApJ 699 (2009) 1153-1170 

\bibitem{Roman-Duval2010} J. Roman-Duval, \etal, ApJ 723 (2010) 492-507 

\bibitem{Ferrand2012} G. Ferrand and S. Safi-Harb, AdSpR 49 (2012) 1313-1319

\bibitem{Pedaletti2013} G. Pedaletti, \etal, A\&A 550 (2013) A123

 \bibitem{Seta1994} M. Seta,et al., 2004, AJ, 127, 1098

\bibitem{Thompson1995} D.J. Thompson, et al., 1995, ApJs, 101, 259-286

\bibitem{Guliani2011} A. Guliani, \etal,  ApJL 742 (2011) L30



\end{thebibliography}
\end{document}